# Fast Computational Ghost Imaging using Unpaired Deep Learning and a Constrained Generative Adversarial Network

Fatemeh Alishahi[1] and Amirhossein Mohajerin-Ariaei[1]

falishah@usc.edu, mohajera@usc.edu

[1]Department of Electrical Engineering, University of Southern California, Los Angeles, CA 90089, USA

**Abstract–** The unpaired training can be the only option available for fast deep learning-based ghost imaging, where obtaining a high signal-to-noise ratio (SNR) image copy of each low SNR ghost image could be practically time-consuming and challenging. This paper explores the capabilities of deep learning to leverage computational ghost imaging when there is a lack of paired training images. The deep learning approach proposed here enables fast ghost imaging through reconstruction of high SNR images from faint and hastily shot ghost images using a constrained Wasserstein generative adversarial network. In the proposed approach, the objective function is regularized to enforce the generation of faithful and relevant high SNR images to the ghost copies. This regularization measures the distance between reconstructed images and the faint ghost images in a low-noise manifold generated by a *shadow* network. The performance of the constrained network is shown to be particularly important for ghost images with low SNR. The proposed pipeline is able to reconstruct high-quality images from the ghost images with SNR values not necessarily equal to the SNR of the training set.

## 1.Introduction

Computational imaging has been highly impacted by the advances in deep learning-based algorithms throughout recent years [1-8]. Particularly, ghost imaging (GI) [9-10] which is an exotic imaging technique can be benefited widely from the potentials offered by deep learning [11-15]. GI can replace the conventional imaging techniques in situations where the object is difficult to reach, or the environment is noisy or harsh for setting-up complex optical systems [16,17]. GI was first introduced within the realm of quantum optics through photon entanglement [18,19] and it has been later verified for classical pseudo-thermal light sources [9-10, 20]. Interestingly, in GI an object is imaged by collecting the light using only a single pixel, "bucket" detector which does not have any resolution. A level of image reconstruction is possible by taking the correlation pattern of the beam scattered from an object and captured by the bucket detector with a reference beam, never seen by the object but spatially correlated with the beam illuminating it. The reference beam pattern at the correlator plate does not need to be detected by a high-resolution detector and there is no need for availability of such beam, since it can be precomputed numerically as in the case for computational ghost imaging (CGI) [11,14,16,17,21,22].

 In practice, to obtain high-resolution images via GI or CGI, it is important to illuminate the object by the pseudo-thermal light for a long period of time to capture enough correlation patterns [12-15,23]. Long exposure time leads to generation of images with sufficient signal-to-noise (SNR) values [23-25]. The number of measurements for high SNR ghost imaging should be much larger than the number of pixels required to resolve an object [11-13]. Aside from the fact that imaging time becomes quite long, this is particularly not ideal for situations involving moving objects or when real time imaging is required.

On the other hand, a not sufficiently long exposure time results into faint ghost images of the object. It has become possible to reconstruct images with better quality from the faint ghost images by employing algorithms such as compressed sensing and Gerchberg–Saxton [26-30] or using deep learning-based ghost imaging (GIDL) as an inverse problem solver [11-15]. The latter method has shown to be superior to the former in terms of higher reconstruction quality for shorter exposure time [12-13]. In other words, GIDL has shown to be able to recover images of an object with a number of measurements lower than the number of pixels needed to resolve the image.

While GIDL can offer an end-to-end and generalized technique through which a variety of images can be reconstructed, still it heavily counts on the availability of a large, paired set of the data which is composed of faint ghost images and their corresponding ground truth copies [11-15]. As discussed earlier, the exploiting of such data is time consuming and cumbersome if not impractical. Furthermore, the requirement for a set of paired images can be particularly difficult in most practical situations where the high SNR images of the object cannot be acquired simultaneously along with the faint ghost images. One practical example of such situations is as follows. The pseudo-thermal statistics of the illuminating light beam can be emulated by modulating of the light wavefront using a spatial light modulator (SLM) [13]. To obtain high SNR ghost images given a limited acquisition time, the SLM pattern should change fast to provide enough number of ensemble random noise patterns. This requires a high-speed SLM which simply may not be available. On the other hand, the high SNR images cannot be obtained through numerical simulations since it may cause spurious artifacts.

Unpaired approaches have been used in the context of unsupervised learning by employing generative adversarial networks (GANs) for translation between different images [30-32] or low-light image enhancement [33]. Inspired by this capability, we propose an end-to-end learning pipeline based on Wasserstein GAN (WGAN) [34,35] which generates high SNR images from low SNR, hastily shot ghost copies without paired supervision. We used WGAN since it has shown to provide smoother objective function through working with a critic network in place of the discriminator network [34,35].

In our method, a maximum posteriori estimation approach [36], which favors images with higher probability conditioned on the faint ghost images is adopted. This results in a regularization term in the generator cost function which quantifies the similarity between the generated images and the faint ghost images. Therefore, the generator has to play a game with the constrain of generating images as similar as possible to its input. We discuss that the effect of this regularized term can be diminished when the ghost images are overly noisy as a result of very short exposure time. Hence, this term cannot dictate an effective similarity constrain between a faint ghost image and its corresponding generated image. To overcome this difficulty, we propose to use a *shadow* generator network which follows the changes of the original generator network. The shadow network transfers the ghost image, and its reconstructed copy produced by the generator to a low-noise manifold. The original generator cost function is then further penalized for generated images which have large distance with the ghost images in the new manifold. Using this proposed architect, ghost images with SNR coefficients between 0.25-1 are retrieved without the access to the exact paired high-quality ground-truth images. Moreover, the pipeline developed here, is shown to be able to reconstruct high-quality images from the ghost images with SNR values different from the SNR of the training set.

## 2. Computational Ghost Imaging

The concept of classical GI is based on the correlation of a pseudo-thermal light beam scattered from an object and captured by a bucket detector with a reference beam [9-10]. Based on this idea a source of a pseudo-thermal light beam is split into two twin waves with a high degree of spatial correlation as shown in Fig.1 (a). One wave is aimed at an object and then captured by a bucket detector while the other part (reference beam) is detected by a pinhole detector [9-10]. Then, the correlation of the two signals detected by the pinhole and bucket detectors is taken to reveal the information about the object. Later it has been proved

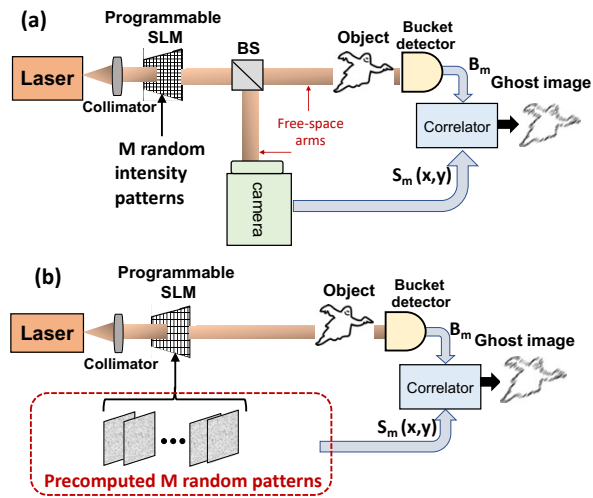

Fig.1 (a): Concept of ghost imaging (GI). (b): Concept of computational ghost imaging (CGI)

that the detection of the reference light by the pinhole detector is not necessary as it can be numerically computed, leading to the emergence of CGI [21,22]. The concept of CGI is plotted in Fig. 1(b). In practice, the pseudo-thermal light is emulated by modulating the beam of a laser using an SLM. The SLM pattern is varied over time and at each time increment $\Delta t$, one random pattern is printed on the SLM. A total number of $M$ patterns can be printed on the SLM for a given amount of time. More accurate images of the object can be obtained after allowing a long enough interaction time or equivalently a large $M$ value.

As mentioned previously, in CGI, the time statistics of the pseudo-thermal light source is pre-computed and can be modeled as the illuminations with a number of random and independent speckle intensity patterns [11,14,15]. The bucket detector integrates the modulated images of the object with the random speckle patterns. The bucket detector outputs the following variable:

$$B_m = \iint O(x,y) S_m(x,y) dx dy \quad (1)$$

in which $x$ and $y$ are the transverse coordinate of the object plane, $O(x,y)$ is a 2D object and $S_m(x,y)$ is the random speckle pattern at time $m\Delta t$. The ghost images are reconstructed by taking the correlation of $B_m$ and the speckle intensity patterns, that is:

$$GI(x,y) = \langle (B_m - \langle B_m \rangle)(S_m - \langle S_m \rangle) \rangle \quad (2)$$

where $\langle . \rangle$ denotes the ensemble average. The SNR of the images is directly related to the exposure time or the number of speckle patterns [25,26]. The SNR is proportional to the SNR coefficient $\beta$, which is equal to $M/N$ in which $M$ is the number of random speckle patterns used for the illumination of object and $N$ is the number of speckles of each random intensity pattern [13,25]. Usually, the value of $N$ is fixed in GCI as speckle patterns are precomputed based on the setup specifications. A larger $\beta$ value means a larger SNR and a higher-quality ghost image. On the other hand, a larger $\beta$ value implies a larger acquisition time which may be practically difficult to achieve. In practice, it is required to reprogram the SLMs with a large number of speckle patterns during a limited acquisition time. This may further complicate the optical setup.

In order to decrease the acquisition time or meet the restriction imposed by the optical system, the capabilities of deep learning architectures have been exploited to recover high SNR images of the object from the ghost images acquired over a low exposure time. In this case, deep learning enables the ghost imaging system to enjoy a much shorter acquisition time.

## 3. Wasserstein GAN for unpaired computational ghost imaging

GANs are employed for unpaired image-to-image translation [30,31]. However, for our application the GAN's ability to generalize the knowledge learnt from a given dataset can come at a price of generated images being irrelevant to its original high SNR version. This problem may overcome by using the concept of cycle-consistency which uses two reciprocal GANs learning the forward and inverse functions [32]. Cycle-consistent GANs are time-consuming to be trained as a result of adding a second trainable network. To avoid the cycle-consistent approach, a regularization term is added to the objective function. This term measures the distance between the generated images and the faint ghost images. In addition, to exploit the functionality of the GAN networks with a more stable and smoother objective function, a refined version of GAN, namely Wasserstein GAN is implemented here. The Mathematical foundation of the developed pipeline is as follows.

Let $x \sim P_X(x)$ and $y \sim P_Y(y)$ denote samples drawn from the original and low SNR ghost images respectively. The problem of reconstructing high SNR images from the low SNR images can be transformed to maximizing a posteriori probability $P_{X|Y}(x|y)$. This can be further converted to finding a high SNR image $x$ that maximizes the following:

$$\text{argmax}_x [\log(P_X(x)) - \lambda_1 \|x - y\|^2] \quad (3)$$

in which $\lambda_1$ is a hyperparameter. Assuming a function $G$ that maps the low SNR images to high SNR images, the above optimization problem becomes

$$\max_G \left[ \mathbb{E}_G \left( \log \left( P_X(G(y)) \right) \right) - \lambda_1 \mathbb{E}_y (\|G(y) - y\|^2) \right] \quad (4)$$

The above equation can be transformed to minimizing a regularized Kullback–Leibler (KL) distance [37] between the probability distributions of generator function, $P_G$, and that of high SNR original images, $P_X$, as:

$$\min_G [KL(P_G, P_X) + \lambda_1 \mathbb{E}_y (\|G(y) - y\|^2)] \quad (5)$$

One can show that the solution to the above optimization problem is identical to the solution of a zero-sum game with the following goal function [34]:

$$\min_G [\max_D JS(D,G) + \lambda_1 \mathbb{E}_Y(\|G(y) - y\|^2)] =$$
$$\min_G \left[ \max_D \left[ \mathbb{E}_x(\log(D(x))) + \mathbb{E}_G \left(1 - \log\left(D(G(y))\right)\right) \right] \right] + \lambda_1 \mathbb{E}_y \min_G (\|G(y) - y\|^2) \quad (6)$$

in which $JS(D,G)$ is the Jensen–Shannon distance [37] between the discriminator, $D$, and generator functions. It should be noted that, aside from the last regularization term, the above equation represents the well-known vanilla GAN objective function in which the two networks of generator and discriminator act in opposite directions to reach to a Nash equilibrium [38]. The problem with vanilla GAN is that for a sub-optimal generator there is a possibility that the gradient of the objective function may shrink, and the generator stops effective learning. Also, the JS divergence in eq. (6) may not be continuous which adds to the difficulties in training step. To remedy such difficulties in training and end up with a GAN architecture that offers a smoother objective function, the WGAN [34,35] was introduced which leverages the Wasserstein distance instead of JS distance [39]. In this new architecture, the discriminator network $D$, now called as the *critic*, has to belong to a k-Lipchitz family of functions [34,35] and provides a continuous score instead of a bounded function. In this regard, the objective function in eq. (6) becomes

$$\min_G \left[ \max_{D \in \mathcal{L}} \left[ \mathbb{E}_X(D(x)) - \mathbb{E}_G\left(D(G(y))\right) \right] \right] + \lambda_1 \min_G [\mathbb{E}_y(\|G(y) - y\|^2)] \quad (7)$$

in which $\mathcal{L}$ denotes family of k-Lipchitz functions. The last term of the above equation is a regularization term that penalizes the objective function for generated images that have a discrepancy with the input ghost images. This term makes the one-path training possible despite the lack of one-to-one relation between the low SNR and high SNR original images. In such, in contrast to cycle consistent GANs, there is no need to train another network that models the inverse problem of generating low SNR images from high SNR images and input the generated images of that network to the original network.

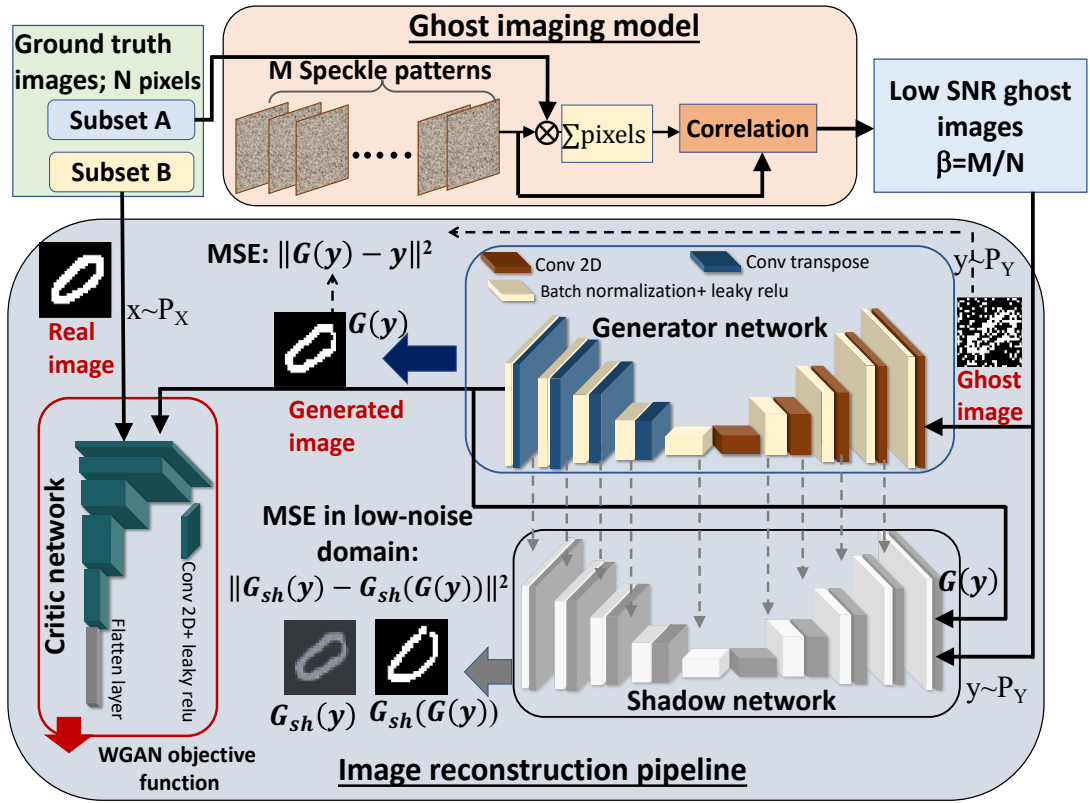

Fig 2. Concept of fast high SNR ghost imaging using an unpaired set of data and a one-path GAN.

## 4. Network Architecture

The concept of fast high SNR ghost imaging using an unpaired set of data and a one-path GAN is plotted in Fig. 2. In the first step, the low SNR ghost images $y \sim P_Y$ are generated numerically from a subset A of original images $x \sim P_X$ as follows: (i) a series of $M$ random intensity speckle patterns representing the SLM masks are multiplied with the original $N$-pixeled image $x \sim P_X$. (ii) each generated intensity pattern is integrated over all its pixels to model the bucket detector output. (iii) the correlation of the random speckle patterns and the output of the bucket detector will be obtained which results in generation of ghost images with an SNR coefficient of $\beta = M/N$.

In the second step, the ghost images $y$ with SNR $\beta$, are fed to the generator network $G$ which plays a zero-sum game with the critic network $D$ as explained in the previous section. The images at the output of the generator $G(y)$, along with the subset B of original images (mutually exclusive with subset A) are loaded into the critic network to produce a critic score. Through the game, the critic network learns to estimate a scoring function that maximizes the Wasserstein distance between the high SNR generated images and the original set of images. Moreover, the generator's goal on one hand is learning to generate high SNR images resembling the original set of images and on the other hand is to generate images as close as possible to its input low SNR ghost copies due to the regularization term. The mean-square-error (MSE) which measures the similarity between the ghost images at the input of the generator network and the generated images at the output of the generator is subtracted from the Wasserstein distance according to eq. (7). This is to penalize the objective function for the images irrelevant to the input ghost images. This latter term can be particularly important for our application. The reason is a good generator network tends to generate plausible images which could have been randomly drawn from the real image dataset. As a result, the high SNR generated images, although good enough to fool the critic, may not have any relevance to the ghost images fed to the generator.

Although the relevancy of the generated images can be preserved to some extent by the regularization MSE term $E_y(\|G(y) - y\|^2)$; through our tests, it has been observed that for lower $M$ values corresponding to lower SNR, the generator can produce high SNR images not related to its noisy input image. To rectify this, we propose to consider another regularization term which measures the similarity between the generated images and their corresponding ghost images after transferring them into a domain with higher SNR. To do so, a shadow network $G_{sh}$, which has a similar structure to the generator network is considered. We consider the following update rule for the shadow network:

$$G_{sh}^i = (1-\alpha)G_{sh}^{i-1} + \alpha G^{i-1} \tag{8}$$

in which $G_{sh}^i$ is the shadow generator at the $i^{th}$ step and $\alpha$ is the update rate. Throughout our experiments $\alpha$ has been kept to 0.1. The generated images and the ghost images are gated to the shadow network to generate $G_{sh}(G(y))$ and $G_{sh}(y)$ respectively. The distance between $G_{sh}(y)$ and $G_{sh}(G(y))$ is expected to show the relevance of the images at the input and the output of the generator. Figure 3 illustrates how this distance can more appropriately indicate the similarity between the generated and ghost images than the distance between $G(y)$ and $y$. In fact, when the noise level is strong, the low SNR ghost images are not distinguishable and therefore, they will have an almost similar and small distance with each other. However, since the shadow network will follow the generator network, it can map the low SNR ghost images to a new manifold where images are more distinguishable or are adequately far apart. This is due to the fact that the generator network has learnt to generate high SNR images through the self-supervised approach of GAN. As shown in Fig. 3, the input domain of shadow network is mapped to a new manifold in which ghost images are separated with larger distances. Therefore, while in the input domain of the shadow network, the ghost images can have similar distances to a particular high SNR image, in the output manifold there is a more unique relation between the distances of the mapped ghost images and their original high SNR copies.

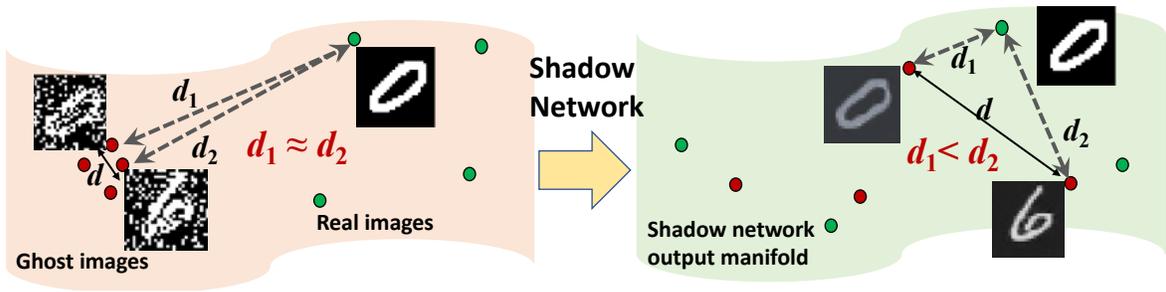

Fig 3. The shadow network maps the ghost images to a manifold in which each image has the smallest distance to its low noise version.

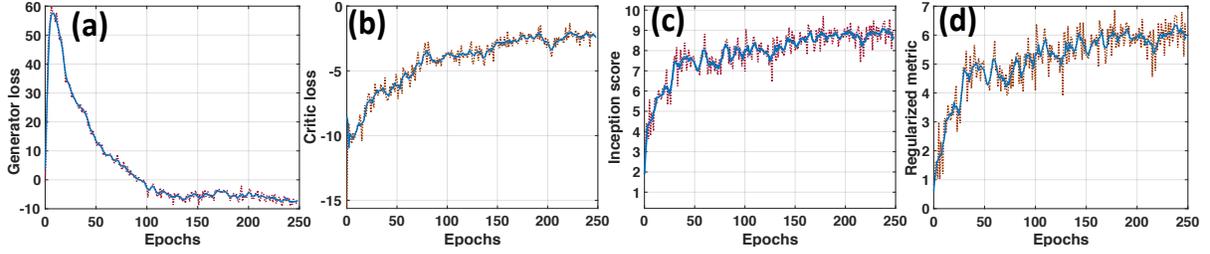

Fig. 4. Dotted lines: (a) generator loss (b) critic loss (c) inception score (d) and the regularized metric with respect to training epochs for ghost images with $\beta$=0.25. Blue solid lines show the filtered versions with a smoothing of 0.6.

Since the distance at the output of the shadow network can better reflects the faithfulness of the generated images, a new MSE term is included in the objective function which penalizes the generator for generated images which are not close enough to original images at the output manifold of the shadow network. The complete WGAN objective function is as follows:

$$L = L_{WGAN} + \lambda_1 MSE_1 + \lambda_2 MSE_2 \quad (9)$$

in which, $L_{WGAN}$ is the WGAN objective function with gradient penalty term [35] and $\lambda_1$ and $\lambda_2$ are hyperparameters. The $MSE_1$ is the regular MSE of the ghost and the generated images or $E_y(\|G(y) - y\|^2)$ and $MSE_2$ is the $l_2$ distance between the transferred ghost and the transferred generated images at the output of the shadow network or $E_y(\|G_{sh}(G(y)) - G_{sh}(y)\|^2)$.

## 5. Training and results

A set of 70,000 handwritten digits of 28 by 28 pixels from MNIST database is chosen and the relevant ghost images of these digits are numerically produced for different $M$ values of 196, 392 and 784 corresponding to SNR coefficients $\beta$, equal to 0.25, 0.5 and 1 respectively. A set of 10,000 images of ghost images is set aside for evaluating the network in the test step. To obtain an unpaired training set of ghost and ground truth images, two mutually exclusive subsets of 30,000 ghost and ground truth images are selected from larger 60,000 sets. In this way, we make sure there is no ghost image of a particular original image present in our training set.

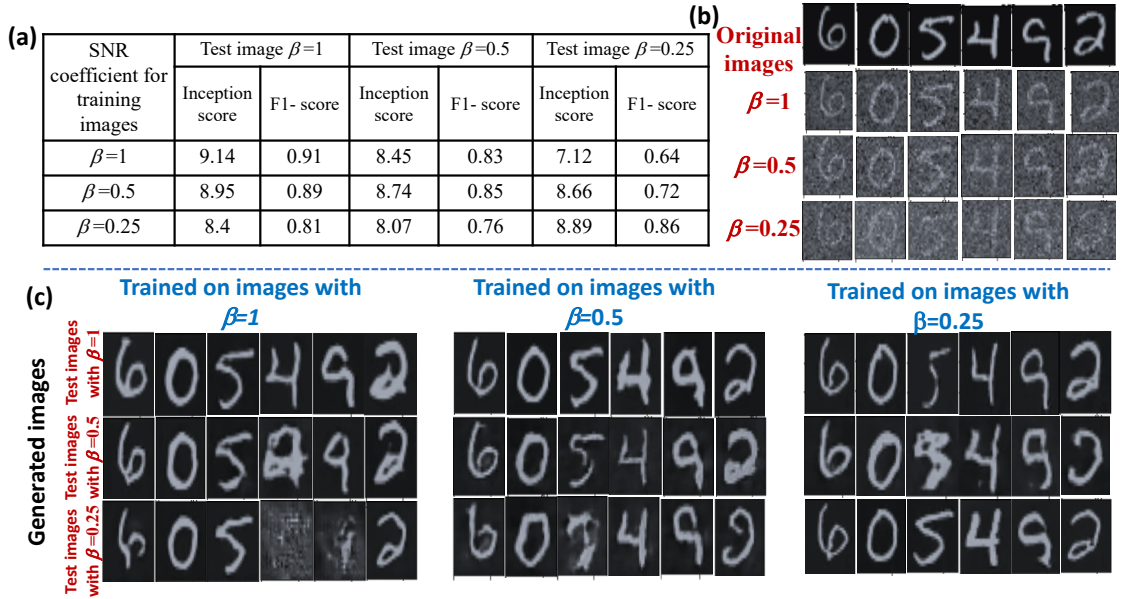

(a)

| SNR coefficient for training images | Test image $\beta$=1 | | Test image $\beta$=0.5 | | Test image $\beta$=0.25 | |
|---|---|---|---|---|---|---|
| | Inception score | F1-score | Inception score | F1-score | Inception score | F1-score |
| $\beta$=1 | 9.14 | 0.91 | 8.45 | 0.83 | 7.12 | 0.64 |
| $\beta$=0.5 | 8.95 | 0.89 | 8.74 | 0.85 | 8.66 | 0.72 |
| $\beta$=0.25 | 8.4 | 0.81 | 8.07 | 0.76 | 8.89 | 0.86 |

Fig. 5. (a) The table of Inception score and macro F1 score for test data of different SNR coefficients of $\beta$ = 25%, 50% and 100% obtained by networks trained on different ghost images with $\beta$ = 25%, 50% and 100%. (b) The ground truth test images and test ghost images with $\beta$ = 25%, 50% and 100%. (c) sample generated images of three networks trained on ghost images with $\beta$ = 25%, 50% and 100% and being tested on data with different SNR levels.

The set of 30,000 ghost images are fed to a generator network which has a conventional encoder-decoder [40] architecture as depicted in Fig. 2. The generator is composed of 5 convolutional blocks consisting of convolutional, batch normalization and LeakyReLU layers for the encoder and 8 de-convolutional blocks consisting of de-convolutional, batch normalization and LeakyReLUs for the decoder part. The critic network is made of 4 convolutional blocks including convolutional and LeakyReLU layers and a flatten layer composed of 2048 neurons.

Although the WGAN distance provides a smooth objective function and is related to the evolution of the critic and the generator networks through the game, it does not speak well to the quality and diversity of the generated images. To address this, a classifier network is pre-trained on MNIST handwritten digits to extract the inception score. The inception score has been proved to better reflect the convergence of GANs working on image data as it is attributed to the ability of the generator to produce varieties of images [41]. Our pre-trained classifier yields an inception score of 9.87 for the original MNIST dataset. In the light of this modification, a customized metric, called regularized inception score is introduced which is defined as:

$$Regularized\ inception\ score = IS - MSE_1 - MSE_2 \qquad (10)$$

in which, $IS$ is the inception score of the generated images. It should be mentioned that $MSE_1$ and $MSE_2$ controls the relevance of the generated images to the input ghost images as discussed earlier. The Adam optimizer with learning rate of 1e-4 and a batch size of 256 is selected. The training process takes around 5 hours on an NVIDIA Tesla P-100 GPU accelerator.

Figure 4 shows different metrics of our GAN during the training of the network on an unpaired set of ground truth and ghost images with $\beta = 0.25$. Obviously, generator and critic losses progress oppositely which embodies the zero-sum gain between the two networks. The inception score and the regularized metric rise with advancement through the epochs which reflects the generation of digits of different classes while maintaining the maximum resemblance to the input ghost images.

Figure 5 shows sample generated images and the summary of different performances for training on multiple sets of ghost images with $\beta = 0.25, 0.5$ and $1$. The performances are evaluated in terms of the inception score and macro F1-score of the

(a) Inception score and macro F1-score average for different test images obtained by networks which are trained on images with $\beta = 0.5$ and use different regularization coefficients generated images from ghost copies with $\beta$ = 0.25, 0.5 and 1. Note that the higher inception score shows the ability of our generator network to produce variety of the classes. On the other hand, the F1-score average quantifies the degree to which these different classes have been correctly generated. As inferred from table in Fig 5(a), the highest inception score and macro F1-score are obtained when the network is evaluated on test images with the same noise level as its training dataset. However, the network is still able to perform well for test images generated from ghost images with different SNR values.

| SNR coefficient for test images | $\lambda_1 = 10$, $\lambda_2 = 10$ | | $\lambda_1 = 20$, $\lambda_2 = 0$ | | $\lambda_1 = 0$, $\lambda_2 = 20$ | | $\lambda_1 = 0$, $\lambda_2 = 0$ | |
|---|---|---|---|---|---|---|---|---|
| | Inception score | F1-score | Inception score | F1-score | Inception score | F1-score | Inception score | F1-score |
| $\beta = 1$ | **8.95** | **0.89** | 8.76 | 0.81 | 8.92 | 0.86 | 9.05 | 0.11 |
| $\beta = 0.5$ | **8.74** | **0.85** | 8.71 | 0.63 | 8.44 | 0.80 | 9.16 | 0.11 |
| $\beta = 0.25$ | **8.66** | **0.72** | 7.8 | 0.51 | 7.7 | 0.68 | 9.1 | 0.13 |

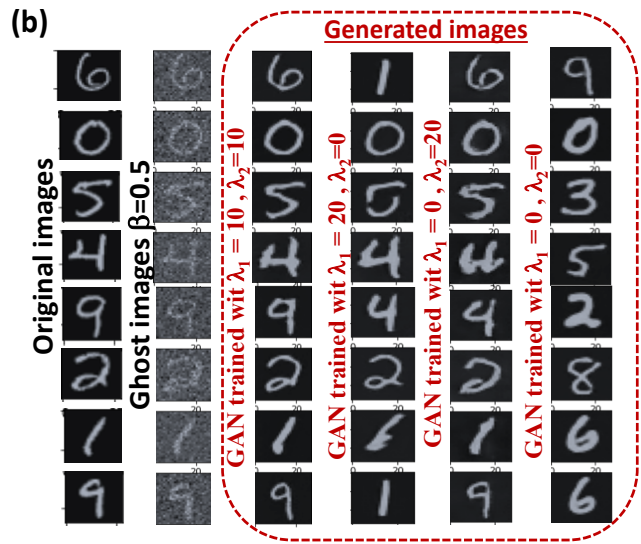

Fig 6. (a) The table of inception score and F1-score average score for test data of different SNR coefficients of $\beta = 0.25, 0.5$ and $1$ obtained by networks trained on ghost images with $\beta = 0.5$ and with different regularization coefficients. (b) The test ghost images with $\beta = 0.5$ and sample generated images of three networks having different values for $\lambda_1$ and $\lambda_2$.

Moreover, it is important to investigate the effect of different regularization terms, $MSE_1$ and $MSE_2$ on the generated images. This is depicted in Fig. 6. A set of ghost images with $\beta$ = 0.5 is chosen for the training set. The network is trained with different values for the regularization coefficients of $\lambda_1$ and $\lambda_2$. The performances of different trainings are assessed by multiple sets of test data having different SNR coefficients. Among different combinations of $\lambda_1$ and $\lambda_2$, the best performance is obtained for $\lambda_1 = \lambda_2 = 10$. It can be seen that, when $\lambda_2 = 0$, the inception score decreases moderately while the macro F1-score shows a more drastic decrease compared to the best result. This is because when $\lambda_2 = 0$, the similarity between generated and ghost images is solely measures in high-noise domain. Therefore, for low SNR ghost images with $\beta$ = 0.5, the generated images do not have high correlation with their ground truth counterparts resulting the macro F1-score to become smaller. However, the inception score can still be relatively high due to the good performance of the GAN network for generation of images belonging to different classes. The ability of the trained GAN to produce variety of images is clear from images generated for $\lambda_1 = \lambda_2 = 0$. As expected, such a trained network shows a very poor F1-score of 11% and a good inception score of ~ 9. The effect of shadow network is evident from the performance achieved with regularization coefficients of $\lambda_1 = 0$ and $\lambda_2 = 20$. In this case, both inception and macro F1-scores decrease from the values obtained for $\lambda_1 = \lambda_2 = 10$. However, the F1-score is decreased by a noticeably smaller amount in this case compared to the case of $\lambda_1 = 20$ and $\lambda_2 = 0$. This is due the fact that $MSE_2$ which measures the fidelity of the generated images uses a distance metric in a lower-noise manifold.

In conclusion, we have developed an end-to-end deep learning pipeline based on a constrained WGAN to reconstruct high-SNR images from low SNR ghost images. The proposed constrained WGAN is trained on two mutually exclusive sets of high SNR and low SNR ghost images and therefore the training is done in an unsupervised fashion on unpaired sets of data. To keep the generated images relevant, two regularization terms are included in the objective function which (1) quantifies the similarity between the reconstructed images and the faint ghost images and (2) measures the distance between the reconstructed images, and the faint ghost images in a low-noise image manifold generated by a non-trainable *shadow* network. High SNR version of ghost images with SNR coefficients of 0.25-1 are reconstructed using the developed pipeline. The individual and mutual effects of the two regularized terms are investigated in terms of the inception score and macro F1-score average of the reconstructed images. It has been found that at SNR coefficients of < 0.5 the effect of the second regularization term is more crucial.